\documentclass[%
 aip,
 amsmath,amssymb,
 reprint,%
]{revtex4-1}

\usepackage{graphicx}% Include figure files
\usepackage{xcolor}
\usepackage{dcolumn}% Align table columns on decimal point
\usepackage{bm}% bold math
\usepackage{comment}

\usepackage[utf8]{inputenc}
\usepackage[T1]{fontenc}
\usepackage{mathptmx}

\begin{document}

   % Math styles
  \def\ind#1{{_{\mathrm{#1}}}}
  \newcommand{\vectornorm}[1]{\left|\left|#1\right|\right|}
  \renewcommand{\vec}[1]{\boldsymbol{#1}}   % \boldsymbol{\mathbf{#1}}
  \newcommand{\uvec}[1]{\hat{\vec{#1}}}
  \renewcommand{\tensor}[1]{\mathbf{#1}}

   % Shorthand
  \newcommand{\figref}[1]{Fig.~\ref{fig:#1}}
  \newcommand{\eqnref}[1]{Eq.~(\ref{eq:#1})}

  % Revisions
  \newcommand{\AS}[1]{\textcolor{red}{ALEX: #1}}
  \newcommand{\SZ}[1]{\textcolor{green}{SEB: #1}}
  \newcommand{\AN}[1]{\textcolor{magenta}{ANA: #1}}
  \newcommand{\JH}[1]{\textcolor{blue}{JENS: #1}}
  \newcommand{\rev}[1]{\textcolor{black}{#1}}

\preprint{AIP/123-QED}

\title[]{\rev{Optimal} motion of triangular magnetocapillary swimmers}
%\title[Magnetocapillary swimmers]{Lattice Boltzmann simulation of triangular magnetocapillary swimmers}
% Force line breaks with \\

\author{Alexander Sukhov}
 \affiliation{Helmholtz Institute Erlangen-N\"{u}rnberg for Renewable Energy (IEK-11), Forschungszentrum J\"{u}lich, F\"{u}rther Stra{\ss}e 248, 90429 N\"{u}rnberg, Germany}

\author{Sebastian Ziegler}%
\affiliation{Institute for Theoretical Physics, Friedrich-Alexander University  Erlangen-N\"{u}rnberg, 91054 Erlangen, Germany}%

\author{Qingguang Xie}
\affiliation{Department of Applied Physics, Eindhoven University of Technology, P.O. box 513, NL-5600MB Eindhoven, The Netherlands}%

\author{Oleg Trosman}
\affiliation{Institute for Theoretical Physics, Friedrich-Alexander University  Erlangen-N\"{u}rnberg, 91054 Erlangen, Germany}%

\author{Jayant Pande}
\affiliation{Department of Physics, Bar-Ilan University, 52900 Ramat Gan, Israel}%

\author{Galien Grosjean}
\affiliation{Universit\'{e} de Li\`{e}ge, GRASP Lab, CESAM Research Unit, All\'{e}e du 6 Ao\^{u}t 19, Li\`{e}ge 4000, Belgium}%

\author{Maxime Hubert}
\affiliation{Universit\'{e} de Li\`{e}ge, GRASP Lab, CESAM Research Unit, All\'{e}e du 6 Ao\^{u}t 19, Li\`{e}ge 4000, Belgium}%

\author{Nicolas Vandewalle}
\affiliation{Universit\'{e} de Li\`{e}ge, GRASP Lab, CESAM Research Unit, All\'{e}e du 6 Ao\^{u}t 19, Li\`{e}ge 4000, Belgium}%

\author{Ana-Sun\u{c}ana Smith}
\affiliation{Institute for Theoretical Physics, Friedrich-Alexander University  Erlangen-N\"{u}rnberg, 91054 Erlangen, Germany}%

\author{Jens Harting}
 \email{j.harting@fz-juelich.de}
 \affiliation{Helmholtz Institute Erlangen-N\"{u}rnberg for Renewable Energy (IEK-11), Forschungszentrum J\"{u}lich, F\"{u}rther Stra{\ss}e 248, 90429 N\"{u}rnberg, Germany}
 \affiliation{Department of Applied Physics, Eindhoven University of Technology, P.O. box 513, NL-5600MB Eindhoven, The Netherlands}%

\date{\today}% It is always \today, today,

\begin{abstract}
A system of ferromagnetic particles trapped at a liquid-liquid interface and subjected to a set of magnetic fields (magnetocapillary swimmers) is studied numerically using a hybrid method combining the pseudopotential lattice Boltzmann method and the discrete element method. After investigating the equilibrium properties of a single, two and three particles at the interface, we demonstrate a controlled motion of the swimmer formed by three particles. It shows a sharp dependence of the average center-of-mass speed on the frequency of the time-dependent external magnetic field. Inspired by experiments on magnetocapillary microswimmers, we interpret \rev{the obtained maxima of the swimmer speed by the optimal frequency centered around the characteristic relaxation time of a spherical particle.} It is also shown that \rev{the frequency corresponding to the maximum speed} grows and the maximum average speed decreases with increasing inter-particle distances at moderate swimmer sizes. The findings of our lattice Boltzmann simulations are supported by bead-spring model calculations.
\end{abstract}

\maketitle

\section{\label{sec:level1}Introduction}

Microswimmers are a paradigmatic example of active matter, and have come into focus in the last two decades with the development of non-equilibrium physics \cite{LaPo09}. Typically, microswimmers are associated with motion at low Reynolds numbers, when self-propulsion dominates over stochastic diffusion, at least in the case of artificial devices. Actually, because self-propulsion requires breaking of the time-reversal symmetry in the stroke, the design of  microswimmers often involves a compromise between engineering a device with the smallest possible degrees of freedom (at least two) and obtaining a time-irreversible stroke \cite{ElWi15}. These difficulties are circumvented in a number of devices \cite{Purc77, AvKe05, DoSt09}, however, the devices based on interacting beads gained particular attention starting with the analysis of a linear arrangement of three spherical beads by Najafi and Golestanian \cite{NaGo04}. In this design, the stroke involves a phase-shift between the contractions and elongations of the two arms connecting the central bead and the external ones, propelling the swimmer along the line \cite{GoAj08}. This model was used to identify some of the fundamental properties of microswimming including the difference between pushers and pullers \cite{PaSm15, DaLi18}, interaction between two swimmers \cite{PoAl07}, and of the swimmer with a wall \cite{ZaNa09, DaLi18}. 

An advantage of the bead-based design is its amenability to experimental investigations. For example, a linear swimmer was achieved employing three ferromagnetic beads placed on a surface of water, which due to the balancing of repulsive magnetic interactions between the induced dipole moments and the capillary attraction, adopt a metastable, linear configuration \cite{GrHu16}. The oscillation of the arms is generated by oscillating magnetic fields, such that the overall force applied to the device remains zero on average \cite{LaGr16}. This design is fundamentally different to the original theoretical counterpart in that the stroke is not imposed but emerges from the balance of forces on the bead, an effect that was concomitantly incorporated into the theoretical modelling and simulations \cite{Feld06, PaSm15, PiPa17, PaMe17,PaMe17a}.

\begin{figure}[htb]
\centering
\includegraphics[width= 0.48\textwidth]{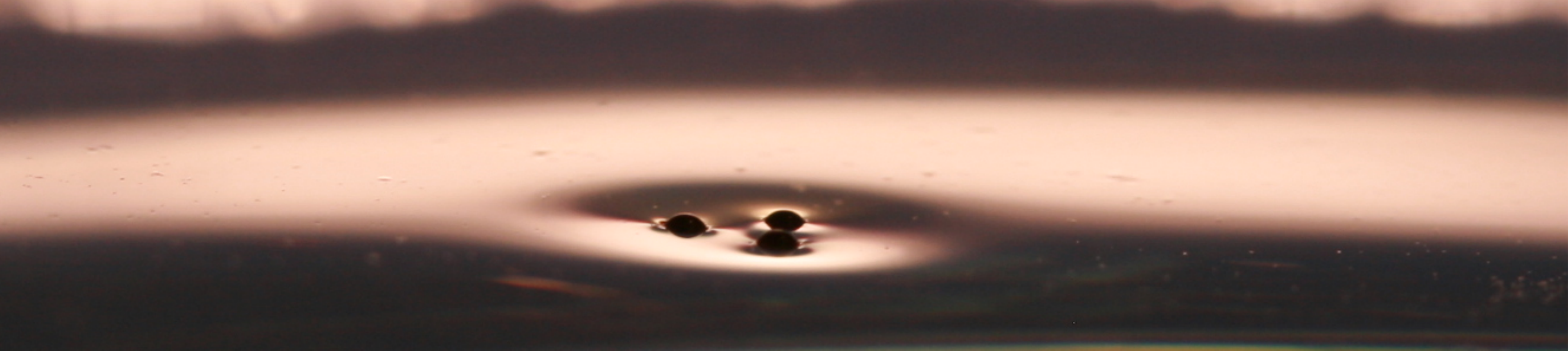}
\caption{Photograph showing the experimental triangular magnetocapillary swimmer in a Petri dish.}
\label{fig:exp_freq}
\end{figure} 

A more stable configuration is a triangular arrangement of magnetic particles at the interface, a real experimental situation of which is demonstrated in Fig.~\ref{fig:exp_freq}\cite{GrLa15}. The particles form an equilateral triangle at an air-water interface. The size of the triangular arrangement is controlled by an external vertical static magnetic field, while the swimmer is driven by a time-dependent magnetic field aligned in the plane of the swimmer. 
%Typically, the average speed of the center of mass 
%of the three particles shows a peak as a function 
%of the frequency of the oscillating magnetic 
%field. 

Numerous efforts involving bead-based devices focused on triangular arrangements that allow for the rotation of the swimmer \cite{RiFa18}, and a richer interaction with external flows \cite{KuLo16}. The first models were concerned with modelling the swimming stroke of Chlamydomonas reinhardtii \cite{FrJu12, PoFr13, BeGo13, JiZi17}. Alternative propulsion mechanisms included modelling a stroke in which the angle between the extendable arms is fixed \cite{LeLo12}, and a stroke in which both arm lengths and the enclosed angle vary \cite{EaPo07}. Further efforts involve a design imposing motion of three beads on a circle \cite{DrBa05} or rotating beads \cite{LoLo08}. In the context of force-based description, internal driving was recently considered based on internal forces acting along the lines connecting the beads arranged in a triangular geometry \cite{RiFa18}. 

Although numerous theoretical studies are known for swimmers in the form of linear \cite{Feld06, ZaNa09, PiGo12, TaMi13, PaSm15, BaLo16} or equilateral arrangements \cite{ChLa15}, we are not aware of any simulations that include a rigorous description of the corresponding hydrodynamic flows, the time dependent deformation of the interface and thus the related capillary effects.
It turns out to be rather challenging to analytically take the partial immersion of the swimmer particles in the liquid, particle rotations, inter-particle forces and their surface effects into account \cite{ChLa15}.
In the present study we aim at bridging this gap by employing the lattice Boltzmann (LB) method for the dynamics of the fluids coupled to a discrete element method for the particle dynamics. 
We perform extensive numerical simulations of triangular magnetocapillary swimmers consisting of three ferromagnetic beads. Our LB-simulations contain the full translational and rotational dynamics of the particles, the dynamics of the fluid flow and can provide insight into the fluid-particle interaction. 
Combining our simulations with a force driven bead spring model of a triangular swimmer, we
explore the relation between propulsion velocity and the frequency of driving. 
%and find a behaviour which can be associated with a mechanical resonance of an overdamped, driven harmonic oscillator.  

The remainder of this article is organized as follows: in the following section we present the numerical method which is followed by a systematic investigation of the equilibrium particle configuration in the presence of capillary and magnetic interactions. Then, we demonstrate our results of the dynamic behaviour of the triangular magnetocapillary swimmer. Subsequently, we analyze our simulation results with the help of a semianalytical model and summarize our findings.

%A similar system was established experimentally using a %magnetocapillary triangular configuration, which %involved an equilateral triangular arrangement of beads %and external driving over an externally defined axis %\cite{GrLa15}. Both these designs allow for steering %the swimmer, albeit using somewhat different %mechanisms.  
 
%This wealth of designs helped in identifying the common features of microswimming, including the quadratic dependence of the propulsion velocity on the driving amplitude \cite{Feld06}, the non-monotonic velocity-viscosity relation \cite{PaMe17}, as well as the non-universal pusher/puller nature of the swimmer \cite{KlFr15}. However, further effort is necessary to identify common features of force-based protocols. 

\section{Simulation method and setup}

\begin{figure}[htb]
\centering
\includegraphics[width= 0.45\textwidth]{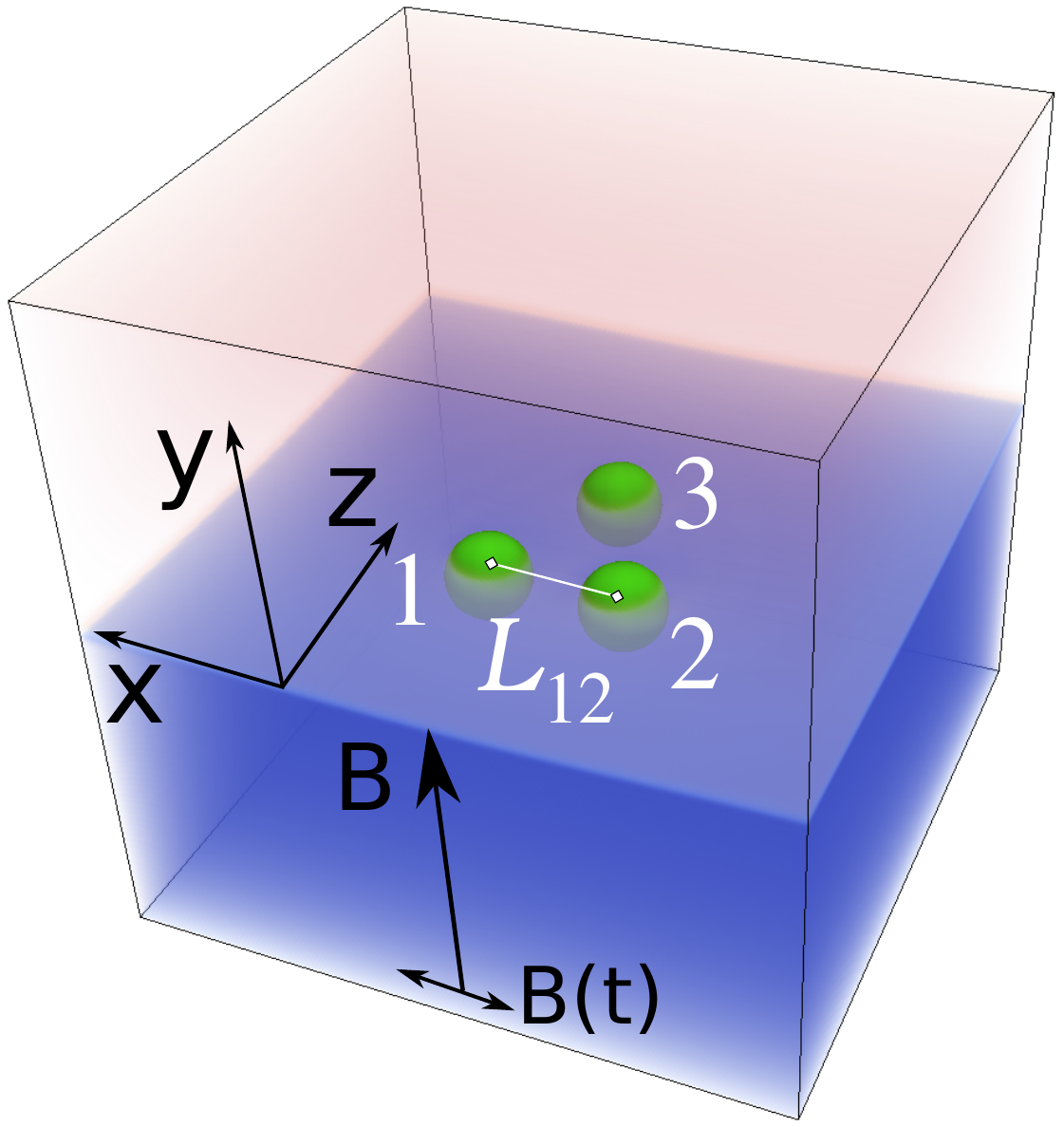}
\caption{Schematics of the simulated system \rev{(box of $128^3$ cubic cells)} including the alignment of external magnetic fields in three dimensions, numbering of beads and inter-particle distances $L_{ij}$ in the plane of the interface.}
\label{fig:geom}
\end{figure} 

For  the  simulation  of  the fluids  we  use  a  lattice  Boltzmann  (LB) method~\cite{Benzi1992}. The LB method allows a straightforward implementation of complex boundary conditions and due to the locality of the algorithm it is well suited for the implementation on parallel supercomputers. It is based on a discretized version of the  Boltzmann  equation
\begin{equation}
    \displaystyle f_i^{c}(\vec{x}+\vec{c}_i\Delta t, t+\Delta t) = f^{c}_i(\vec{x},t)+\Omega^{c}_i(\vec{x},t), 
\label{eq:LBE}    
\end{equation}
which describes the time evolution of a single-particle distribution function $f^{c}_i(\vec{x},t)$ at time $t$ and position $\vec{x}$. $\vec{c}_i$ denotes the discrete velocity vector in the $i$th direction for fluid component $c=\{1, 2\}$. Here, we use a so-called D3Q19 lattice with $i=1, \ldots, 19$\cite{Qian1992}. The left hand side of Eq. (\ref{eq:LBE}) describes the free streaming of fluid particles, while their collisions are modelled by a Bhatnagar-Gross-Krook (BGK) collision operator on the right hand side as \cite{BhGr54}
\begin{equation}
    \displaystyle \Omega^{c}_i(\vec{x},t) = - \frac{f^{c}_i(\vec{x},t)-f^{\mathrm{eq}}_i(\rho^{c}(\vec{x},t),\vec{u}^{c}(\vec{x},t))}{\tau^{c}/\Delta t}. 
\label{eq:BGK}
\end{equation}
Here, $f^{\mathrm{eq}}_i(\rho^{c}(\vec{x},t),\vec{u}^{c}(\vec{x},t))$ is a third-order
%\JH{?} \AS{Consulted with Qingguang. In the "\texttt{input-file}" the parameter "\texttt{bdist}" is responsible for this function. 
%In all my calculations for the swimmer, this parameter "\texttt{bdist=2}", meaning that it was \underline{a third order the distribution function}.} 
equilibrium distribution function, and macroscopic densities and velocities are given by $\rho^{c}(\vec{x},t)=\rho_0\sum_i f^{c}_i(\vec{x},t)$ and $\vec{u}^c(\vec{x},t)=\sum_i f^c_i(\vec{x},t)\vec{c}_i/\rho^c(\vec{x},t)$, respectively ($\rho_0$ is a reference density). $\tau\ind{c}$ is the relaxation rate of component $c$, which determines the relaxation of $f^{c}_i(\vec{x},t)$ towards the equilibrium. We employ a three-dimensional lattice with the cell size $\Delta x$ in space and the time $t$ is discretized with $\Delta t$-steps. The speed of sound $c\ind{s}=1/\sqrt{3}\Delta x / \Delta t$ depends on the choice of the lattice geometry and allows one to obtain the kinematic $\nu^{c}=c^2_{\mathrm{s}}\Delta t (\tau^{c}/\Delta t - 1/2)$ or the dynamic $\eta^{c}=\nu^{c}\rho^{c}$ fluid viscosities. For simplicity, we set $\Delta x=\Delta t=\rho_0=\tau^{c}=1$ in the remainder of this paper.

Capillary interactions can be modeled when more than one fluid species is present. In this case a mean-field interaction force between several fluid components is calculated according to the pseudopotential method of Shan and Chen as\cite{ShCh93,Liu2016} 
\begin{equation}
    \displaystyle \vec{F}^c_{\mathrm{C}}(\vec{x},t) = - \psi^c(\vec{x},t)\sum_{c'}g_{cc'}\sum_{\vec{x}'}\psi^{c'}(\vec{x}',t)(\vec{x}'-\vec{x}).
\label{eq:SC}
\end{equation}
Here, $c$ and $c'$ refer to different fluid components, $\vec{x}'$ denotes the nearest neighbours of the lattice site $\vec{x}$ and $g_{cc'}$ describes a coupling constant determining the surface tension. $\psi^c(\vec{x},t)$ has the functional form $\psi^c(\vec{x},t)\equiv \psi^c(\rho^c(\vec{x},t))=1-\mathrm{e}^{-\rho^c(\vec{x},t)}$. The force (\ref{eq:SC}) is applied to the fluid component $c$ by adding a shift $\Delta \vec{u}^c(\vec{x},t)=\tau^c\vec{F}^c_{\mathrm{C}}(\vec{x},t)/\rho^c(\vec{x},t)$ to the velocity $\vec{u}^c(\vec{x},t)$ in the equilibrium distribution. The method is a diffuse interface method, with an interface width of typically $5$ lattice sites depending weakly on the coupling strength \cite{FrGu12, KrFr13}. In the binary fluid system we refer to the fluids as ``red'' (r) and ``blue'' (b) \cite{JaHa11}.

Rigid particles are simulated by solving Newton's equations of motion for translational and rotational degrees of freedom by means of a leap-frog algorithm. The particles are discretized on the lattice and are coupled to both fluid species by means of a modified bounce-back boundary condition for multiple fluid components~\cite{Ladd94,LaVe01,JaHa11,GFH14}. 
%When two particles approach each other very closely so that no %lattice node is available between their surfaces to describe the %hydrodynamic forces, a lubrication correction is applied %\cite{LaVe01}.
\rev{
When a lattice site $\vec{x}$ is occupied by the surface of a particle, the following equation is applied to its neighbouring fluid lattice site $(\vec{x} + \vec{c}_i$)
\begin{equation}
  f_{i}^{c}(\vec{x} + \vec{c}_i , t + 1) = f_{\bar{i}}^{c}(\vec{x} 
  + \vec{c}_i , t) + \Omega_{\bar{i}}^{c}(\vec{x} + \vec{c}_i , t) + C
  \mbox{,}
  \label{eq:momentum-fluid}
\end{equation}
where $C$ is a linear function of the local velocity of particle surface~\cite{JaHa11} and $\bar{i}$ is defined such that $\vec{c}_i=-\vec{c}_{\bar{i}}$.
%\eqnref{momentum-fluid} introduces changing of momentum of fluid. 
In order to conserve the total momentum of the system, 
an additional force $\vec{F}_p$ and torque $ \vec{D}_p$ is applied on the particle~\cite{JaHa11} to 
compensate for the momentum change of the fluid caused by~\eqnref{momentum-fluid}
\begin{eqnarray}
 \vec{F}_p &=& (2f_{\bar{i}}^{c}(\vec{x} + \vec{c}_i , t)+C)\vec{c}_{\bar{i}} \mbox{,} \\
 \vec{D}_p &=& \vec{F}_p \times \vec{r}(t) \mbox{,}
\end{eqnarray}
in which $\vec{r}(t)$ is a vector directed from the particle center to the lattice site of reflection.
While the particle moves, the configuration of lattice sites occupied by the particle changes. For newly occupied sites, the fluid on that site is removed and its momentum is added to the particle through a force~\cite{JaHa11}
\begin{equation}
 \vec{F}_{pn} = -\sum_c \rho^c(\vec{x}, t) \vec{u}^c(\vec{x}, t) \mbox{.}
\end{equation}
Vacated lattice sites need to be filled with fluid. 
In the case of two fluid components, we define an average density~\cite{JaHa11, FrGu12}
\begin{equation}
 \bar{\rho}^c(\vec{x}, t) = \frac{1}{N_{FN}}\sum_{i_{FN}}\rho^c(\vec{x}+\vec{c}_{i_{FN}}, t)
 \mbox{,}
\end{equation}
where $N_{FN}$ is the number of neighbouring fluid sites with coordinates $\vec{x}_{i_{FN}} = \vec{x}+\vec{c}_{i_{FN}}$.
The fluid on the vacated site is then initialised with distribution functions~\cite{JaHa11, FrGu12}
\begin{equation}
 f_{i}^{c}(\vec{x}, t) = \rho_{\mathrm{new}}^{c} \cdot f_{i}^{eq}(\vec{u}_{\mathrm{surface}}(\vec{x}, t), \rho_{\mathrm{new}}(\vec{x}, t))
\mbox{,}
 \end{equation}
where $\vec{u}_{\mathrm{surface}}(\vec{x}, t)$ is the local velocity of the particle surface. 
$\rho^{c}_{\mathrm{new}}(\vec{x}, t)$ corresponds to $\bar{\rho}^c(\vec{x}, t)$ plus a small correction term to account for local density gradients imposed by the Shan-Chen forcing\cite{JaHa11, FrGu12}.
}

%The non-zero repulsive force between the particle surface and the surrounding fluid might introduce a slightly smaller effective fluid density close to the particle surface than the bulk density, which leads to a mass drift over time if one chooses $\rho^{c}_{\mathrm{new}}(\mathbf{x}, t)=\bar{\rho}^c(\mathbf{x}, t)$. 
%To conserve the total mass on long time scales, $\rho^{c}_{\mathrm{new}}(\mathbf{x}, t)$  is chosen to be~\cite{JaHa11, FrGu12}
%\begin{equation}
% \rho^{c}_{\mathrm{new}}(\mathbf{x}, t) = \bar{\rho}^c(\mathbf{x}, t)\left( %1-C_0\frac{\sum_c \rho^{c}_{\mathrm{init}}\Delta %\rho^c(t)}{\rho^{c}_{\mathrm{init}}V_{\mathrm{box}}} \right)
%\end{equation}
%where $\Delta \rho^c(t)$ is the total mass error of fluid $c$ at
%time $t$, $\rho^{c}_{\mathrm{init}}$ is the initial density of fluid $c$,  %$V_{\mathrm{box}}$ is the volume of the system and $C_0=2500$ measures the strength of %the mass corrections. 
\rev{
When two particles approach each other very closely so that no lattice node is available between their surfaces to describe the hydrodynamic forces, a lubrication correction can be applied~\cite{LaVe01}. However, we always assure a sufficient resolution and the particles never get closer than a few $\Delta x$ in the current paper. Therefore, such a correction is not required here.
}

The ferromagnetic particles are considered as rigid spheres with fixed orientation of the magnetic moment with respect to the particle. When applying a static magnetic field $B_{\mathrm{y}}$ along the positive $y$-direction (Fig.~\ref{fig:geom}), repulsive dipolar forces are induced. The magnetic repulsion is balanced by an attractive capillary force which is due to the interface deformation caused by the gravity-induced immersion of the particles. This leads to a stable arrangement of the beads after a certain relaxation time. In analogy with the experiments on magnetocapillary swimmers \cite{LuOb13, GrLa15} we choose the amplitude of the time-dependent magnetic field to be approximately three times lower than that of the static field to treat it as a modulation. The field $\vec{B}(t)=B_{0 \mathrm{x}}\cos \omega t \vec{e}_{\mathrm{x}}$ drives the system out of the local equilibrium leading to a directed collective motion of the beads. 

In our implementation we assume a homogeneous external magnetic field $\vec{B}$ and prescribe a magnetic moment $\vec{\mu}_i=\chi V \vec{B}/\mu_0$ to each particle $i$, where $\chi$ is the particle susceptibility, $V$ is its volume and $\mu_0 = 4\pi \times 10^{-7}$ (in lattice units) corresponds to the magnetic permeability of vacuum. The emergent magnetic dipole-dipole interaction between a pair of particles is then 
\begin{equation}
 \displaystyle U_{ij} = -\frac{\mu_0}{4\pi r_{ij}^3}\left[ 3(\vec{\mu}_i\cdot \vec{e}_{ij})(\vec{\mu}_j\cdot \vec{e}_{ij})-(\vec{\mu}_i\cdot \vec{\mu}_j)\right].
\label{eq_1}
\end{equation}
Here, $r_{ij} \equiv \vectornorm{\vec{r}_{ij}} \equiv \vectornorm{ \vec{r}_i - \vec{r}_j}$ is the distance between the centres of two spheres $i,j$ located at $\vec{r}_i$ and $\vec{r}_j$, respectively, and $\vec{e}_{ij}$ = $(\vec{r}_{i}-\vec{r}_{j})/\vectornorm{\vec{r}_{i}-\vec{r}_{j}}$. 
The magnetic field generated by the magnetic moment $\vec{\mu}_j$ at the location of another particle $i$ is
\begin{equation}
 \displaystyle  \vec{B}_i= - \frac{\delta U_{ij}}{\delta \vec{\mu}_i} = \frac{\mu_0}{4\pi r^3_{ij}}\left[3\vec{e}_{ji}(\vec{\mu}_j\cdot \vec{e}_{ji})-\vec{\mu}_j\right].
\label{eq_1a}
\end{equation}
The resulting magnetic force acting on the $i$th particle is then $\vec{F}_i= - \vec{\nabla} \left(- \vec{\mu}_i\cdot (\vec{B}_i+\vec{B})\right)$, or more explicitly 
\begin{eqnarray}
\displaystyle 
 \vec{F}_{i} =  \frac{3\mu_0}{4\pi r^4_{ij}}  && 
\left( \vec{\mu}_i \left( \vec{\mu}_j \cdot \vec{e}_{ji} \right) 
+ \vec{\mu}_j \left(\vec{\mu}_i \cdot \vec{e}_{ji} \right) \right.  \nonumber \\
&& \left. - 5 \vec{e}_{ji} \left(\vec{\mu}_j \cdot \vec{e}_{ji} \right) \left(\vec{\mu}_i \cdot \vec{e}_{ji} \right) +\vec{e}_{ji}(\vec{\mu}_i \cdot \vec{\mu}_j)\right).
\label{eq_1b}
\end{eqnarray}
Similarly, the magnetic torque acting on the particle $i$ is $\vec{T}_i=\left[\vec{\mu}_i \times (\vec{B}_i+\vec{B})\right]$, or explicitly 
\begin{equation}
\vec{T}_{i}\!\! = \!\!\frac{\mu_0}{4\pi r_{ij}^3}\cdot \left(3\left(\vec{\mu}_j \cdot \vec{e}_{ji}\right)\left[\vec{\mu}_j\!\times\! \vec{e}_{ji} \right] - \left[\vec{\mu}_i \!\times\! \vec{\mu}_j\right]\right) + \left[\vec{\mu}_i\! \times\! \vec{B}\right].\!\!\!
\label{eq_1c}
\end{equation}
The total force and the total torque for each particle include a summation of expressions (\ref{eq_1b}) and (\ref{eq_1c}) over index $j$. We note that the homogeneous external magnetic field $\vec{B}$ does not produce a net force on the particles and serves exclusively to change the direction of the particle magnetic moment.
The magnetic forces and the torques given by Eqs. (\ref{eq_1b}) and (\ref{eq_1c}) are chosen such that they balance capillary forces in the equilibrium.
The method with implemented magnetic interactions has already been benchmarked and successfully applied for simulations of magnetocapillary phenomena \cite{XiDa15, XiDa16, XiDa17}. 

We consider a simulation box consisting of $128^3$ cubic cells containing two equally-sized fluid lamellae (\figref{geom}). Walls with mid-grid bounce back boundary conditions are placed parallel to the interface at $y=0$ and $y=128$, whereas for the boundaries in all other directions we assume periodicity. All particles have equal radii $R=5$ which assures that they are small on the scale of the simulation box, but sufficiently large with respect to the interface thickness\cite{XiDa15, XiDa16, XiDa17}. The beads are initially placed near the interface between the red and blue fluids with densities $\rho_r=\rho_b=0.7$, deforming thus the interface of the fluids due to the presence of a body force mimicking gravity acting on the particles. This leads to attractive capillary forces between the beads. The fluid-fluid coupling constant is chosen $g_{cc'}=0.1$ giving rise to the numerical surface tension $\gamma \approx 0.04$ \rev{in LB-units} (see Ref. \cite{FrGu12})\rev{, which assures a well-defined interface profile on the one hand, and the numerical stability of the method on the other hand}. In all simulations presented here, the two fluids are considered identical, i.e. with equal viscosity and density. Due to the limitations of the pseudopotential lattice Boltzmann method, we are not able \rev{to reach realistic values for the surface tension and density ratios to mimic the air and water phases of the experiment. Even though we keep the dimensionless parameters such as the Bond number close to the experimental values wherever possible, we do expect in particular the limitation in the choice of surface tensions to have an impact on the swimmer propulsion. The ratio of surface tension forces and magnetic forces 
%, i.e. $\frac{2\pi \gamma R}{(\mu_0\mu^2/(4\pi r^4_{pp}))}$, 
for parameters used in the experiment is several orders of magnitude higher than in the simulations. 
In addition, even density ratios of the order of 10 or 20 comprise the numerical stability resulting in a smaller available range of driving frequencies and Bond numbers.
The long-range nature of capillary interactions would require system sizes of at least $10$ times the size of the swimmer in order to remove finite size effects from the simulations~\cite{XiDa16,KHV10,KH11}. However, due to the inherently slow motion of the swimmer, our simulations require millions of time steps and thus the computational time required would be prohibitive. 
When checking the impact of the finite size effects on the maximum swimmer speed, we do find only a factor two increase in the maximum swimming speed even for ten times larger systems and we do not expect the qualitative behaviour of the swimmer to change substantially.  
Thus, we expect quantitative differences in the propulsion of the experimental swimmer as compared to our simulations and therefore refrain from a quantitative comparison. For a quantitative comparison, an improved multiphase model would be required.}

\section{Equilibrium position of magnetic particles at a fluid-fluid interface}

\subsection{Vertical equilibrium for a single particle}

\begin{figure}[!t]
\centering
\includegraphics[width= 0.48\textwidth]{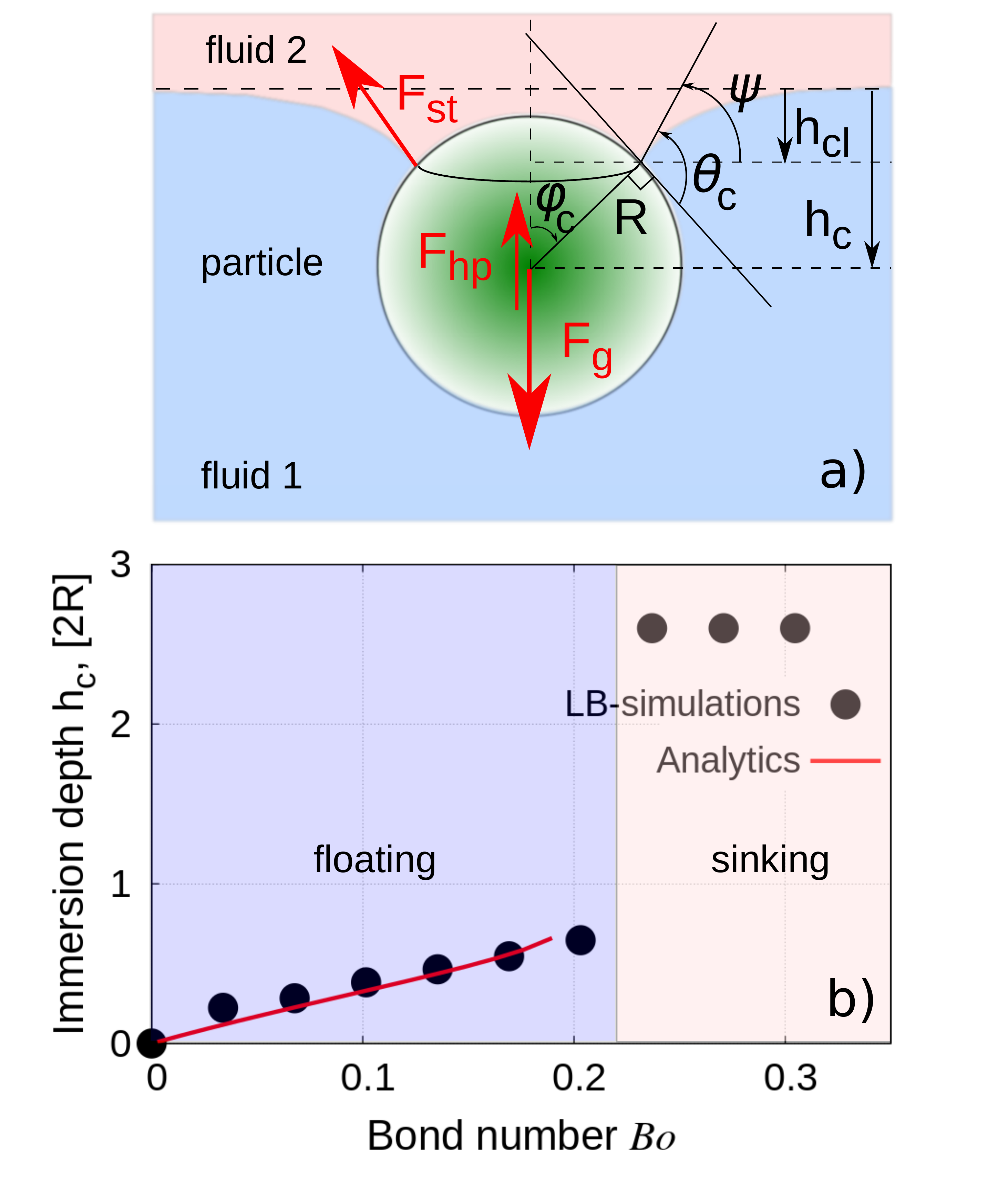}
\caption{a) Schematics of a particle floating at a fluid-fluid interface. b) Immersion depth $h\ind{c}$, measured as a distance of the particle center of mass from the position of the undisturbed fluid-fluid interface, as a function of the Bond-number ($Bo$). LB-simulations are shown by black solid circles and the analytical solution based on Eq. (\ref{eq_3a}) is represented by the solid red curve.}
%$\theta\ind{c}$, which is fixed in the present simulations ($\theta\ind{c}=\pi/2$), is the angle (\textit{contact angle}) between the tangential to the fluid interface and the perpendicular from the particle center of mass, both in the contact point. $\varphi\ind{c}$ is defined as the angle between the vertical line and the line from the center of mass to the contact point. $h\ind{cl}$ is the height of the curved fluid relative to the contact line and $\psi = \theta\ind{c} - \varphi\ind{c}$.
\label{fig:1_bead}
\end{figure} 

Prior to the simulation of multiple particles, we focus first on the equilibrium properties of a single particle at the interface. In the case of a single bead the homogeneous magnetic field exerts no magnetic force. The particle reaches its equilibrium and floats when the gravitational force $\vec{F}\ind{g}$ is balanced by the sum of the surface tension force $\vec{F}\ind{st}$ (or its vertical projection) and the force related to the hydrostatic pressure $\vec{F}\ind{hp}$ (Fig.~\ref{fig:1_bead}a), i.e. $F_{\mathrm{g}}^{\mathrm{y}}=F_{\mathrm{st}}^{\mathrm{y}}+F_{\mathrm{hp}}^{\mathrm{y}}$. More explicitly, the force balance perpendicular to the interface reads \cite{Bink06,VeMa05}
\begin{equation}
 \displaystyle 
 \begin{split}
 \rho\ind{p}g & \frac{4}{3}\pi R^3  = 2 \pi \gamma R \sin \varphi\ind{c} \sin \psi + \pi R^3 (\rho\ind{fl1}-\rho\ind{fl2})g\times \\  
 & \left(\frac{2}{3}+\cos\varphi\ind{c}-\frac{1}{3}\cos^3\varphi\ind{c}+\frac{h\ind{cl}}{R}\sin^2\varphi\ind{c}\right) + \rho\ind{fl2}g\frac{4}{3}\pi R^3, 
 \end{split}
\label{eq_2}
\end{equation}
where $g$ denotes the gravitational constant and $\rho\ind{p}, \rho\ind{fl1}, \rho\ind{fl2}$ are the densities of the particle and the fluids 1 and 2, respectively. The \textit{contact angle} $\theta\ind{c}$, which is fixed in the present simulations ($\theta\ind{c}=\pi/2$), is the angle between the tangent to the fluid interface and the perpendicular from the particle center of mass, both in the contact point. $\varphi\ind{c}$ is defined as the angle between the vertical line and the line from the center of mass to the contact point. Finally, the angle $\psi$ is the difference $\psi = \theta\ind{c} - \varphi\ind{c}$.  $h\ind{c}$ is the immersion depth of the particle and $h\ind{cl}$ is the height of the perturbed liquid relative to the contact line (Fig.~\ref{fig:1_bead}a). 

Based on Eq. (\ref{eq_2}) it is straightforward to introduce a dimensionless quantity, the \textit{Bond number},  which relates the downward- and upward forces
\begin{equation}
 \displaystyle Bo = \frac{(\rho\ind{p}-\rho\ind{fl1})gR^2}{\gamma},
\label{eq_3}
\end{equation}
where it is assumed that the interface is flat ($h\ind{cl}\rightarrow 0$) and the particle is almost completely immersed in the fluid ($\varphi\ind{c}\approx 0$).  Given the parameters of the experimental situation from Refs. \cite{LuOb13, GrLa15}, i.e. particles with a radius of $R^{\mathrm{exp}}=0.25\cdot 10^{-3}$~m, a density of $\rho\ind{p}=7.8\cdot 10^3$~kg/m$^3$ ($\rho\ind{g}\approx 0$) and the surface tension of water $\gamma=73$~mN/m, we obtain for the experimental Bond numbers $Bo^{\mathrm{exp}}\approx 0.06$.

In the present LB-simulations for equal densities  of the red and blue fluids $\rho\ind{fl1}=\rho\ind{fl2}=0.7$ one obtains $Bo^{\mathrm{LB}}=(\rho\ind{p}-\rho\ind{fl2})gR^2/\gamma$. The dependence of $h\ind{c}(Bo)$ is shown in Fig.~\ref{fig:1_bead}b. It is linear up to the point where the particle detaches from the interface.
The LB-simulations are compared with the solution of the equation for the angle $\varphi\ind{c}$ obtained from Eq. (\ref{eq_2}) for $h\ind{cl}\approx 0$ as
\begin{equation}
 \displaystyle \cos^3\varphi\ind{c} -3\cos\varphi\ind{c}-\frac{3}{2k_0 Bo^{\mathrm{LB}}}\sin(2\varphi\ind{c})+\frac{4}{k_0}-2=0,
\label{eq_3a}
\end{equation}
where $k_0=\frac{(\rho\ind{fl1}-\rho\ind{fl2})}{(\rho\ind{p}-\rho\ind{fl2})}$ (note that at the interface $\rho\ind{fl1}-\rho\ind{fl2}=0.04\rho_0$) and a solution of the form $h\ind{c}\approx 2R\cos\varphi\ind{c}$ is found. As inferred from Fig. \ref{fig:1_bead}b the LB-simulations are in good agreement with the solution (\ref{eq_3a}).

In the general case the Bond number for the detachment of the particle from the interface is given by $Bo_{\mathrm{crit}}\approx 3/2\sin^2(\theta\ind{c}/2)|_{\theta\ind{c}=\pi/2}\approx 0.75$\cite{VeLe06}. Corrected by the factor due to the presence of the other fluid, the critical Bond number in the simulations is in line with the analytical prediction $Bo^{\mathrm{LB}}_{\mathrm{crit}}\approx 0.21$.  

\subsection{Equilibrium of two particles}

\begin{figure}[htbp]
\centering
\includegraphics[width= 0.42\textwidth]{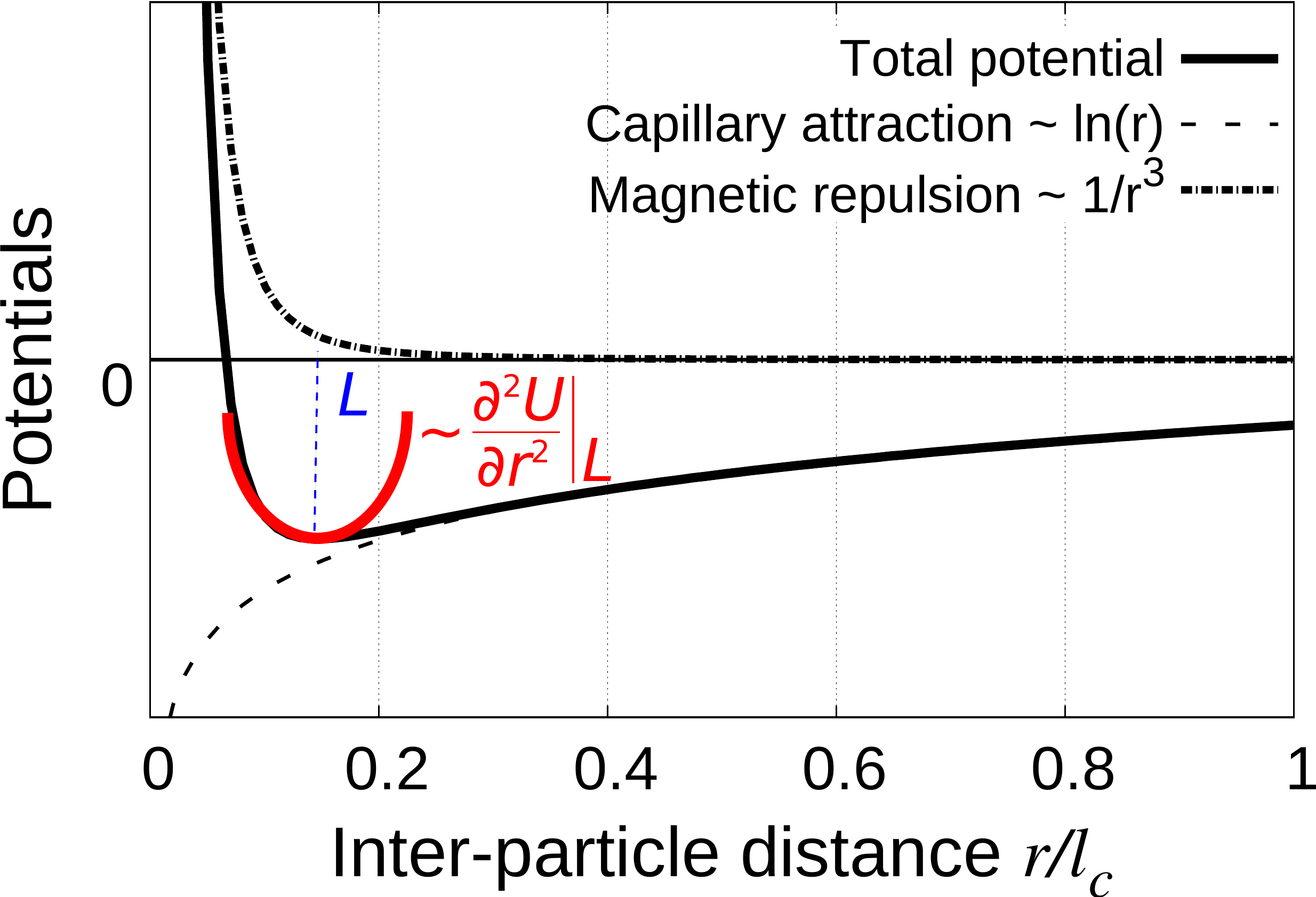}
\caption{Schematics of the magnetocapillary potential given by Eq. (\ref{eq_4}). The first derivative $\partial U^{\mathrm{tot}}/\partial r=0$ defines the equilibrium position $L$ for the energy minimum, whereas the second derivative $\partial^2 U^{\mathrm{tot}}/\partial r^2\equiv k$ determines the strength of the potential around the equilibrium $L$.}
\label{fig:Utot}
\end{figure} 

When two ferromagnetic particles are placed at the fluid/fluid interface at zero magnetic field, they experience a capillary attraction due to the deformation of the interface. For a pair of small particles the capillary potential reads  $U^{\mathrm{cap}}=-C^{\mathrm{cap}}K_0\left(r/l\ind{c}\right)$ (Ref. \cite{VeMa05}), where $r$ is the distance between the centers of the particles, $K_0(x)$ is the modified Bessel function of the second kind, $l\ind{c}=\sqrt{\frac{\gamma}{(\rho\ind{fl1}-\rho\ind{fl2})g}}$ is the capillary length and $C^{\mathrm{cap}}=2\pi \gamma R^2 (Bo^{\mathrm{LB}})^2 \sigma^2$, with $\sigma=\frac{1}{3}(\frac{\rho\ind{p}}{\rho\ind{p}-\rho\ind{fl2}}+1)$ \cite{VeMa05}. Since the capillary length is dependent on the Bond number, i.e. $l\ind{c}=R\sqrt{\frac{(\rho\ind{p}-\rho\ind{fl2})}{(\rho\ind{fl1}-\rho\ind{fl2})}\frac{1}{Bo^{\mathrm{LB}}}}$, we estimate the \textit{minimal} capillary length for a high $Bo^{\mathrm{LB}}=0.16$ in the simulations as $\frac{l\ind{c}}{2R}\approx 7$. In contrast, for a low value of $Bo^{\mathrm{LB}}=0.016$ we obtain for $\frac{l\ind{c}}{2R}\approx 22$.  

In the situation when the inter-particle distance $r$ is much shorter than the capillary length $r \ll l\ind{c}$, the capillary potential can be well approximated by $U^{\mathrm{cap}}\approx C^{\mathrm{cap}}\ln \left( r/l\ind{c}\right)$. For reaching an equilibrium a uniform static magnetic field aligned perpendicularly to the interface is required, since it turns the dipole-dipole interaction to the repulsive one (Eq. (\ref{eq_1}), $U_{12}(\uparrow \uparrow)=\mu_0 \mu_i \mu_j/(4\pi r^3_{ij})$)\rev{, which opposes the attractive capillary interaction for reaching a finite equilibrium distance}. The total potential thus has the form
\begin{equation}
 \displaystyle U^{\mathrm{tot}}(r\ll l\ind{c})=C^{\mathrm{cap}}\ln \left(\frac{r-a}{l\ind{c}}\right) + \frac{C^{\mathrm{mag}}}{(r-a)^3},
\label{eq_4}
\end{equation}
where the term $(r-a)$ accounts for the finite size of the beads, $a=2R$ is the minimal distance between the centres of the beads and $C^{\mathrm{mag}}=\mu_0 \mu_i \mu_j/(4\pi)$. Potential (\ref{eq_4}) is schematically illustrated in Fig.~\ref{fig:Utot} and clarifies the role of the first and second derivatives.

After minimization of Eq. (\ref{eq_4}) one obtains for the equilibrium inter-particle distance
\begin{equation}
 \begin{split}
 \displaystyle \frac{L_{12}}{2R} (r\ll l\ind{c})  = & 1 + \frac{1}{2R}\sqrt[3]{\frac{3C^{\mathrm{mag}}}{C^{\mathrm{cap}}}} = \\
 & 1 + \frac{1}{2}\sqrt[3]{\frac{\left(\frac{3\mu_0 \mu^2_{\mathrm{S}}}{4\pi R^4}\right)}{(2\pi\gamma R) (Bo^{\mathrm{LB}})^2 \sigma^2}}\left(\frac{\mu_i}{\mu\ind{s}}\right)^{2/3},
 \end{split}
\label{eq_5}
\end{equation}
where we note that the ratio $\left(\frac{3\mu_0 \mu^2_{\mathrm{S}}}{4\pi R^4}\right)/(2\pi\gamma R)$ is the ratio of magnetic to capillary forces at equilibrium and \rev{should be of order one; $\mu\ind{S}$ denotes here the saturated magnetic moment}. The equilibrium position for inter-particle distances well below the capillary length thus scales as
\begin{equation}
 \displaystyle \frac{L_{12}}{2R} (r\ll l\ind{c}) = 1 + \frac{1}{2}\left(Bo^{\mathrm{LB}}\sigma \right)^{-2/3}\left(\frac{\mu_i}{\mu\ind{s}}\right)^{2/3}.
\label{eq_6}
\end{equation}

When $r$ approaches the capillary length ($r\approx l\ind{c}$), the logarithmic approximation for the Bessel function no longer holds. A function which approximates the Bessel function in this range very well is $\sqrt{\pi/2}\mathrm{e}^{-r/l\ind{c}}/\sqrt{r/l\ind{c}}$ \rev{(Ref. \cite{BrSe98})}. However, an analytical handling of such a function is complex. An alternative approach is to use an empirical expression $U^{\mathrm{cap}}\approx - 0.45 \frac{C^{\mathrm{cap}}l\ind{c}}{r-a}$ which fits numerically the Bessel function in this range well and allows to write the total potential as
\begin{equation}
 \displaystyle U^{\mathrm{tot}}(r\approx l\ind{c})=-0.45\frac{l\ind{c}C^{\mathrm{cap}}}{r-a} + \frac{C^{\mathrm{mag}}}{(r-a)^3}.
\label{eq_7}
\end{equation}
Minimization of expression (\ref{eq_7}) under similar assumptions as of Eq. (\ref{eq_5}) yields for the equilibrium $L_{12}$ 
\begin{equation}
 \begin{split}
 \displaystyle \frac{L_{12}}{2R} (r\approx l\ind{c}) = 1 + 0.68\frac{\sqrt[4]{\rho\ind{b}-\rho\ind{r}}}{2\rho\ind{p}-\rho\ind{r}} & \left(\rho\ind{p}-\rho\ind{r}\right)^{3/4} \times \\
 \displaystyle & \left(Bo^{\mathrm{LB}}\right)^{-3/2} \left(\frac{\mu_i}{\mu\ind{s}}\right).
 \end{split}
\label{eq_8}
\end{equation}

\begin{figure}[htbp]
\centering
\includegraphics[width= 0.45\textwidth]{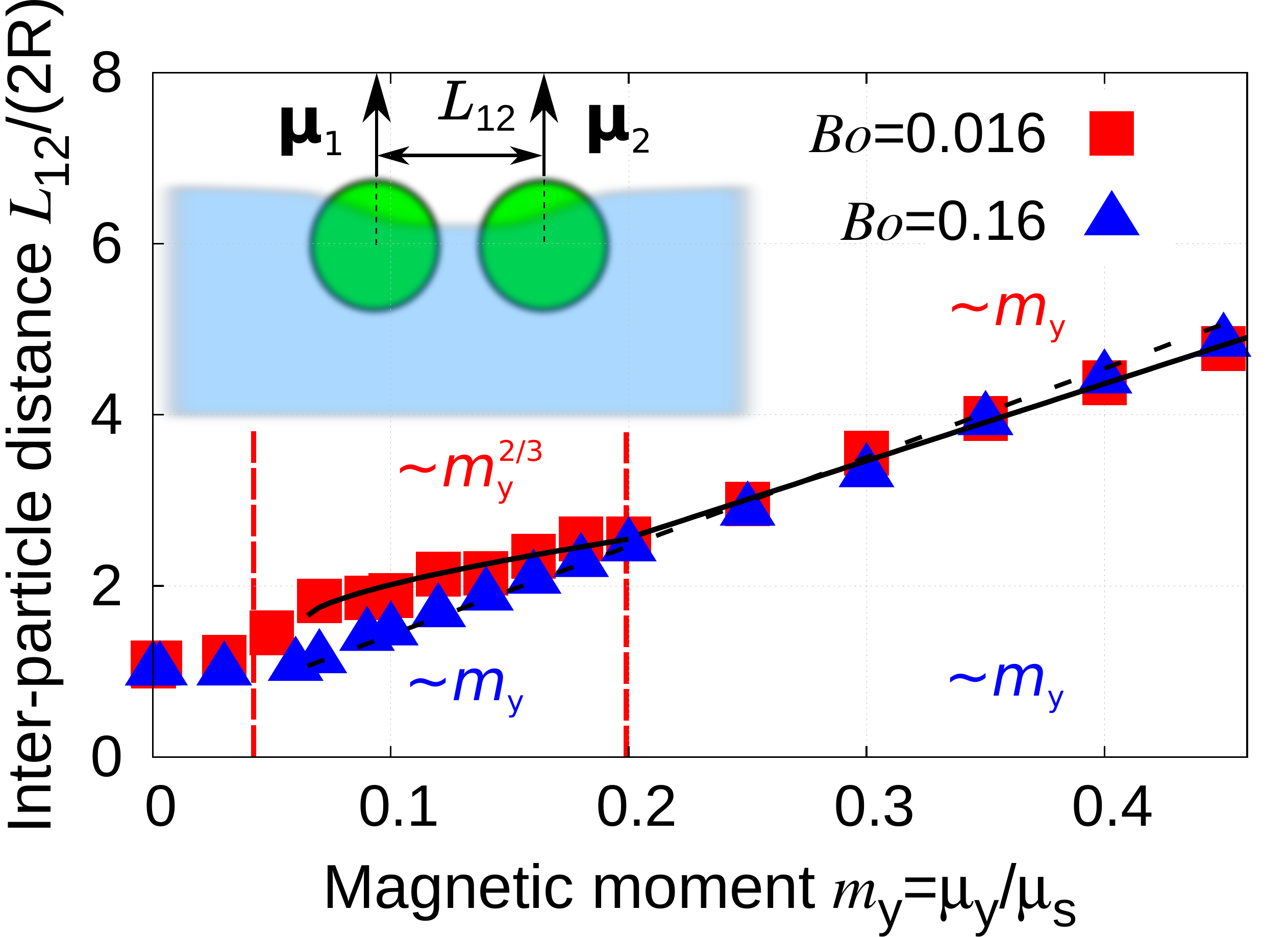}
\caption{Dependence of the equilibrium inter-particle distance $L_{12}$ on the values of the induced magnetic moment for two particles and for various $Bo$-numbers. The results of LB-simulations at equilibrium are shown by solid symbols, the solid curve indicates two areas scaling as $\sim m^{\mathrm{2/3}}_{\mathrm{y}}$ (given by Eq. (\ref{eq_6})) and $\sim m_{\mathrm{y}}$ (given by Eq. (\ref{eq_8})). The dashed line represents a linear $\sim m\ind{y}$-scaling at high $Bo$-number.}
\label{fig:2_beads}
\end{figure} 

Fig.~\ref{fig:2_beads} demonstrates three regimes of the equilibrium $L_{12}$-dependence on the strength of the static magnetic field or the reduced magnetic moment $m_y=\mu\ind{y}/\mu\ind{S}$. The three areas correspond to experimental observations shown in Fig. 3 of Ref. \cite{VaCl12}. Below $m\ind{y}\approx 0.05$ the equilibrium inter-particle distance becomes "locked". This phenomenon is due to \textit{capillary bridges} \cite{KrNa01, MyBa87}, where a small amount of liquid causes sticking of particles together because of the minimized liquid interface. 
%\JH{The force leading to the sticking of two particles %is known for the case when the liquid film is only %between the particles \cite{GoBr10}, while the presence %of the fluid/fluid-interface obviously reduces the %overall force.} 
%\AS{Here I mean that the effect of capillary bridges is %not well studied when there is an interface between %particles. However, the presence of the interface %should make the effect weaker.} 
In the range $0.05<m\ind{y}<0.2$ we observe different growth dependencies for low and high Bond numbers, since the particles are already separated and $L_{12}$ is well below $l\ind{c}$ for $Bo^{\mathrm{LB}}=0.016$ (Eq. (\ref{eq_6})) or approximately around $l\ind{c}$ for $Bo^{\mathrm{LB}}=0.16$ (Eq. \ref{eq_8}). Finally, $L_{12}$ grows linearly for high magnetic moments which is related to the proximity of $L_{12}$ to the capillary length, as given by Eq. (\ref{eq_8}).

\subsection{Equilibrium of three particles}

\begin{figure}[htbp]
\centering
\includegraphics[width= 0.45\textwidth]{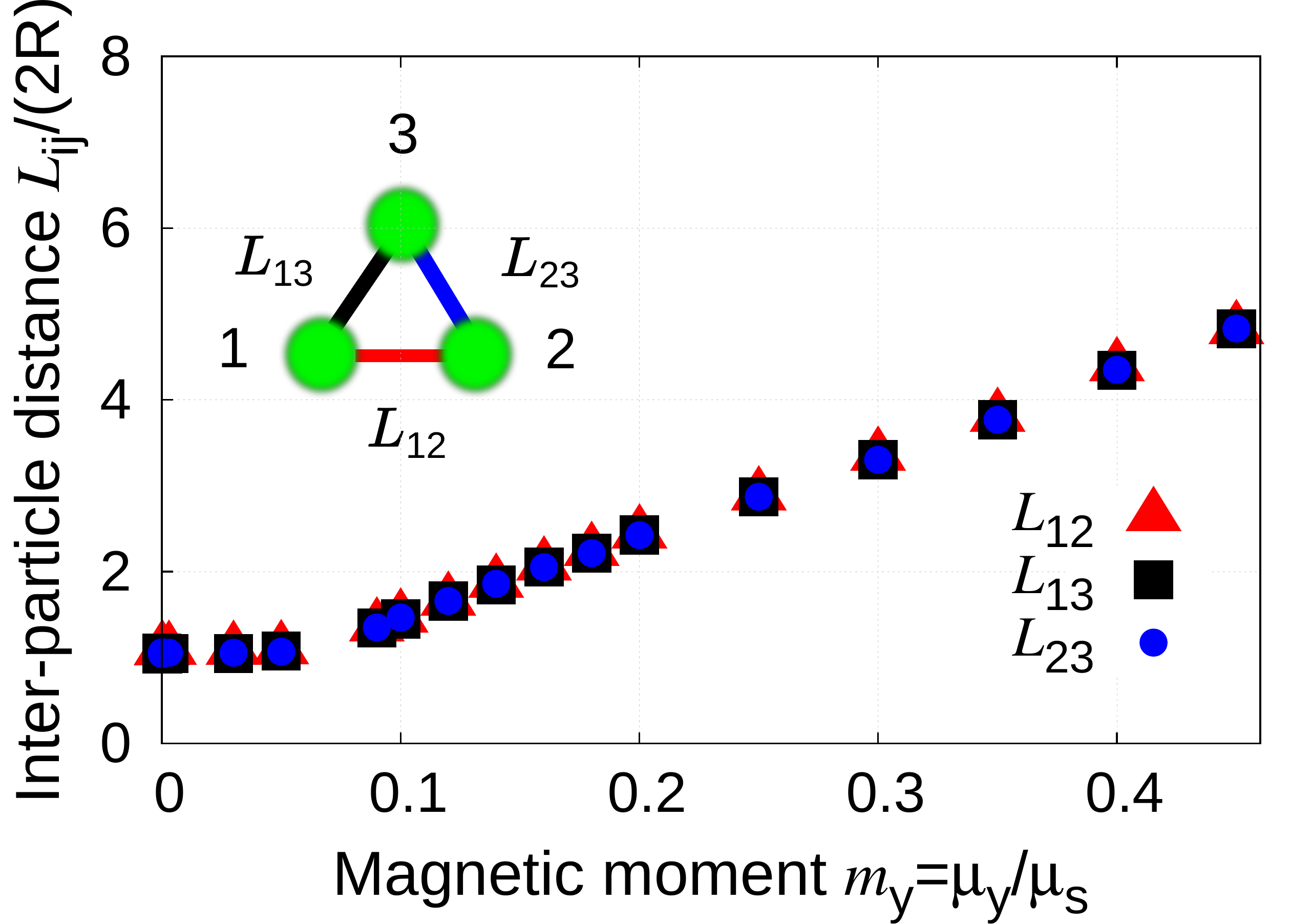}
\caption{Dependence of the equilibrium inter-particle distance $L_{ij}$ on the values of the induced magnetic moment for three particles and the fixed Bond number $Bo^{\mathrm{LB}}=0.16$. }
\label{fig:3_beads}
\end{figure} 

At last we investigate the case of three particles. All three particles are placed onto the interface simultaneously and a vertical static magnetic field (no time-dependent field so far) is applied from the very beginning. The initial inter-particle distance is set to $\frac{L_{ij}}{2R}(t=0)=2.6$. From symmetry considerations we expect an equilateral triangle in the plane of the interface for the equilibrium configuration. As evidenced by Fig. \ref{fig:3_beads}, the length of the sides of the triangle $L_{ij}$ remains equal irrespective of the induced magnetic moment. 

An analytical treatment of the problem is more involved, since it requires taking into account the asymmetry of the capillary potential and the finite size of all three particles. Nevertheless, one can clearly observe the "locked" range until $m\ind{y}\approx 0.05$ and the range of the linear growth for $L_{ij}$.  As in the case of two particles in Fig.~\ref{fig:2_beads}, the present simulations fully recover the measurements demonstrated in Fig. 6 or Ref. \cite{VaCl12}
%\AS{Black dots in Fig. 6 of Ref. [51] are measurements, red curves - simulations; we refer to the measured data}.
Noteworthy is also the fact that the values of the magnetic moment for which the transition from the ``locked'' to the linear regimes takes place, remain quantitatively unchanged for two and three beads at high $Bo=0.16$ (cf. Figs. \ref{fig:2_beads} and \ref{fig:3_beads}). The latter we explain as a compensation of the amount to which the capillary interactions grow by the increased repulsive magnetic interaction all due to the presence of the third particle.

%We finally note that the definition for the \textit{magnetic Bond number} $Bo\ind{m}\equiv (r^{\mathrm{eq}}/(2R))^4$ \cite{VaCl12} is not used here, since in this case it would depend on the distance and moreover the question of the exact definition - whether according to Eq. (\ref{eq_5}) or to Eq. (\ref{eq_7}) - remains open. 

\section{Dynamics of a magnetocapillary swimmer}

\begin{figure}[htbp]
\centering
\includegraphics[width= 0.45\textwidth]{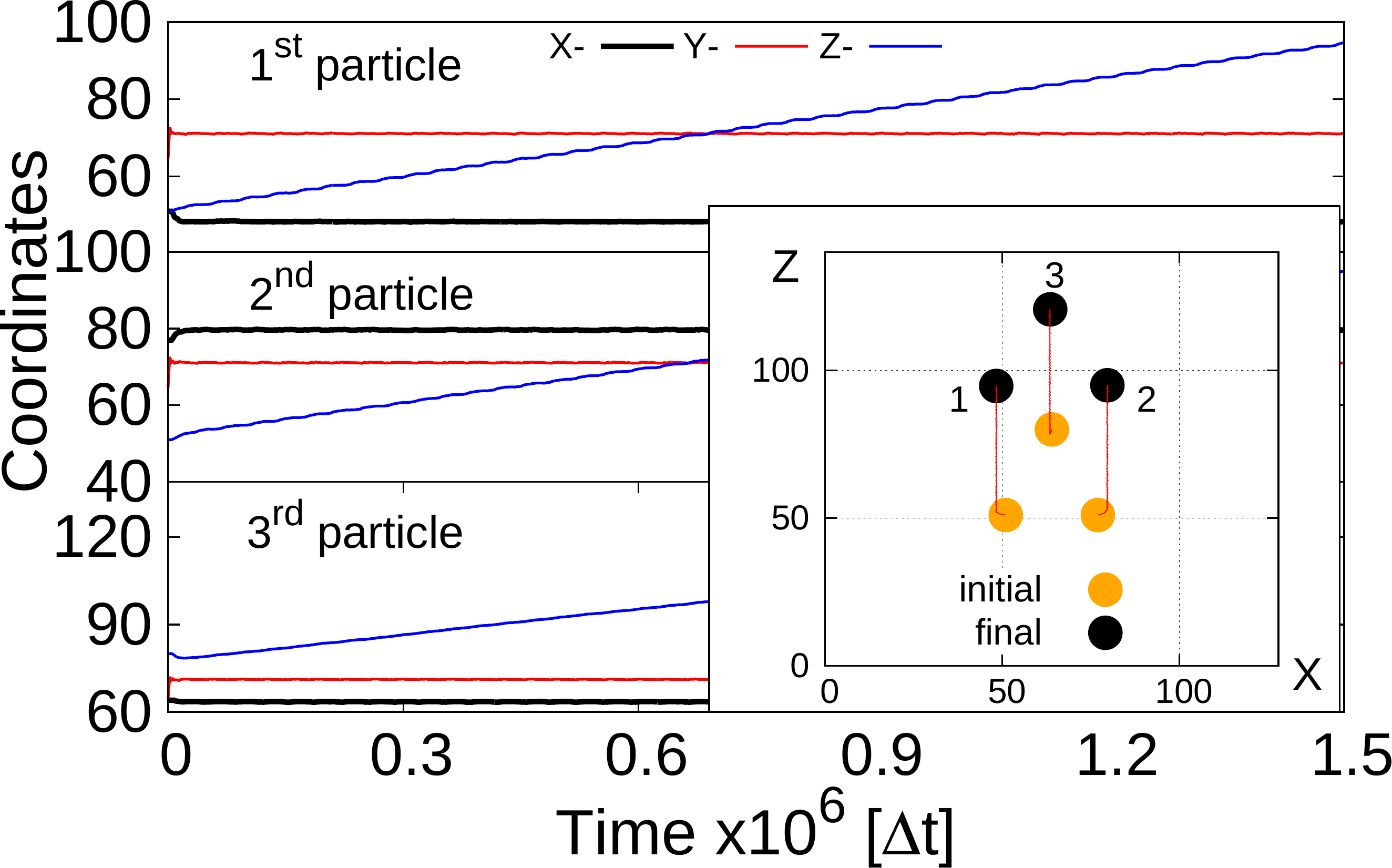}
\caption{Trajectories of each bead during the swimmer motion. The inset shows the initial and final positions of the swimmer on the interface. The frequency of the applied B-field is \rev{$\omega/(2\pi)=0.76$ $(\tau_{\mathrm{cs}})^{-1}$, where $\tau_{\mathrm{cs}}=95$}  (Eq. (\ref{eq:tauv})) and the field ratio $|B(t)|/|B|\approx 0.36$.}
\label{fig:traj}
\end{figure} 

\subsection{Directed motion}

So far, only static configurations formed by one, two or three beads were considered. Now we intend to investigate the dynamical behaviour of the three-particle ensemble that performs directed motion powered by an external time-dependent magnetic field. The ensemble is termed a \textit{magnetocapillary swimmer}.

The simplest stable system which can break the symmetry required by the scallop theorem is a three-beads swimmer having a triangular shape (\figref{geom})\cite{Purc77}. It takes approximately $30000$ time steps for the $x$-coordinate of all beads to reach the stable configuration (cf. onset of the trajectories in \figref{traj}). \rev{This time is chosen rather empirically after examining the relaxation dynamics of particles in the LB-simulations. It assures a full vertical and horizontal equilibrium for the particles and the interface.} Only after this initial relaxation the oscillating time-dependent in-plane magnetic field $\vec{B}(t)=B_{0 \mathrm{x}}\cos \omega t \vec{e}_{\mathrm{x}}$ is applied. For the amplitudes $|B(t)|/|B|\approx 0.36$ we choose a ratio which is close to the one used in the experiments \cite{LuOb13, GrLa15}. In order to stay well below the critical Bond number \rev{for the detachment of the particle from the interface} and to still assure a significant interface deformation we keep $Bo=0.16$. 

For characterization of the swimmer motion we introduce an average velocity of the center of mass \rev{in the non-moving frame} defined by
\begin{equation}
\displaystyle \left<\vec{v}\right>=\frac{1}{3} \sum_{i=1}^3 \frac{\vec{r}_i(t+n2\pi/\omega)-\vec{r}_i(t)}{n2\pi/\omega} = \frac{1}{3}\sum_{i=1}^3 \vec{v}_i, 
\label{eq:v}
\end{equation}
where $n$ counts the number of the periods of the oscillating field and $\vec{r}_i$ stands for the coordinates of the beads' centers.

\figref{traj} demonstrates the motion of the swimmer in the case where the direction of the oscillating field $\vec{B}(t)$ is fixed along the $X$-axis with the fixed given frequency. In this configuration the swimmer propagates itself with minor deviations of all beads from their equilibrium positions and shows a linear motion along the $z$-axis. Careful analysis of all coordinate components reveals that not only do the $Z$-coordinates demonstrate visible oscillations, but also the $Y$-components, i.e. the vertical coordinates, oscillate with amplitudes close to $0.08 R$. \rev{Finally, since the AC magnetic field is applied along the line connecting particles 1 and 2, and since particle 3 is symmetric to both the field and those particles, one can observe stronger oscillations of the $Z$-coordinates for beads 1 and 2.}
%The fact that the $Z$-coordinates of the first and the second particle oscillate stronger than the $Z$-coordinate of the third particle points additionally to the direction of the pumping magnetic field.
\begin{figure}[htbp]
\centering
\includegraphics[width= 0.45\textwidth]{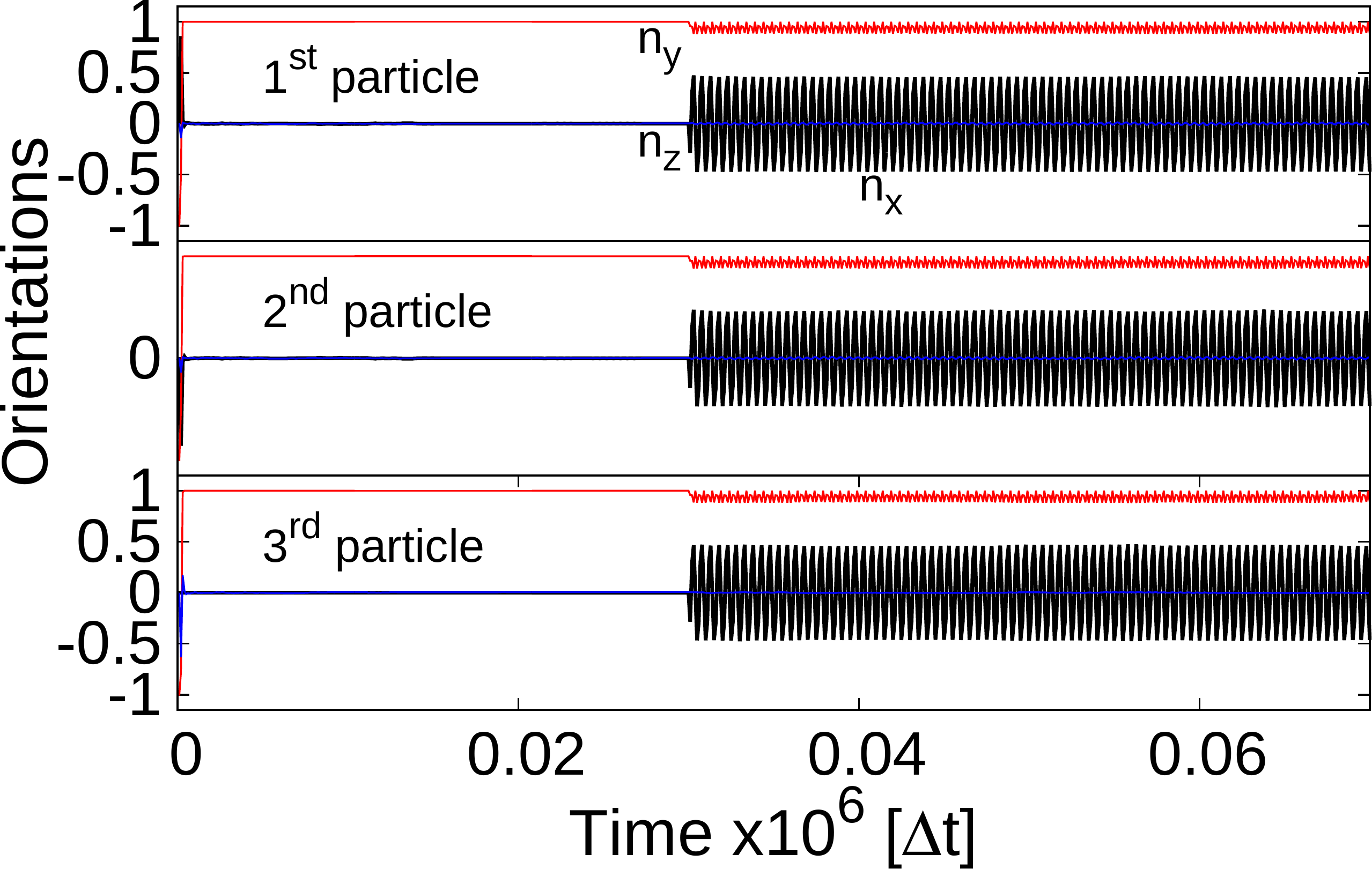}
\caption{Trajectories of the orientation vector $\vec{n}_i$ for each bead on the short time scale. The parameters of the simulations are identical with those for Fig. \ref{fig:traj}.}
\label{fig:orient_rot}
\end{figure} 

Fig.~\ref{fig:traj} does not answer the question \rev{how the particles propagate during the stroke cycle.} This issue is clarified in Fig.~\ref{fig:orient_rot}, where the propagation of the orientation vector (director) $\vec{n}_i$, \rev{coinciding with the magnetization orientation} for each bead is demonstrated. As expected, the particles do not spin until the external oscillating magnetic field starts driving them (after $t=30000$), while upon switching the $\vec{B}(t)$-field on, all three particles start oscillating. \rev{Note, that the particles do not fully rotate around their own axes and the plane of the director oscillations remains always perpendicular to the net swimmer motion}. Although the amplitude of the oscillating field is of the order of one third of the static field amplitude $|B(t)|/|B|\approx 0.36$, the magnetic forces and torques acting on the beads scale as $|F^{\mathrm{magn}}|\sim |B^2|$, resulting therefore in $|F^{\mathrm{magn}}(t)|/|F^{\mathrm{magn}}|\approx 0.13$. This explains why the change in the rotational orientation of all the beads is so weak ($n\ind{y}> n\ind{x}, n\ind{z}$). 

We should point to one significant difference in the propagation mechanism of the swimmer in the experiments\cite{VaCl12, LuOb13, GrLa15, GrHu16, LaGr16} and the implementation presented above. \rev{In the aforementioned experiments a set of three magnetic fields was applied: a strong static field perpendicularly to the interface to induce the magnetic repulsion between the beads, an in-plane oscillating magnetic field to modify the local equilibrium and a weak static in-plane field applied under an angle to the oscillating one to select a unique oscillation mode\cite{GrLa15}. The action of the three fields led to a strong rotation of the particles \textit{in the plane of the interface}. The magnetic forces and the torques implemented using Eqs. (\ref{eq_1b}) and (\ref{eq_1c}), respectively, are not supposed to result in any in-plane rotations of the particles and demonstrate small out-of plane oscillations of the beads (Fig. \ref{fig:orient_rot}).} For this reason we expect different speeds of propulsion within the present simulations and the experiments.

%While in the experiments a set of several magnetic fields is applied, leading to a strong rotation of the particles \textit{in the plane of the interface}, the magnetic forces and the torques implemented using Eqs. (\ref{eq_1b}) and (\ref{eq_1c}), respectively, are not supposed to result in any in-plane rotations of the particles and should rather demonstrate small out-of plane oscillations of the beads (Fig. \ref{fig:orient_rot}). For this reason we expect different speeds of propulsion within the present simulations and the experiments.

\subsection{Frequency dependence of the swimmer motion}
\begin{figure*}[htb]
\centering
\includegraphics[width= 0.9\textwidth]{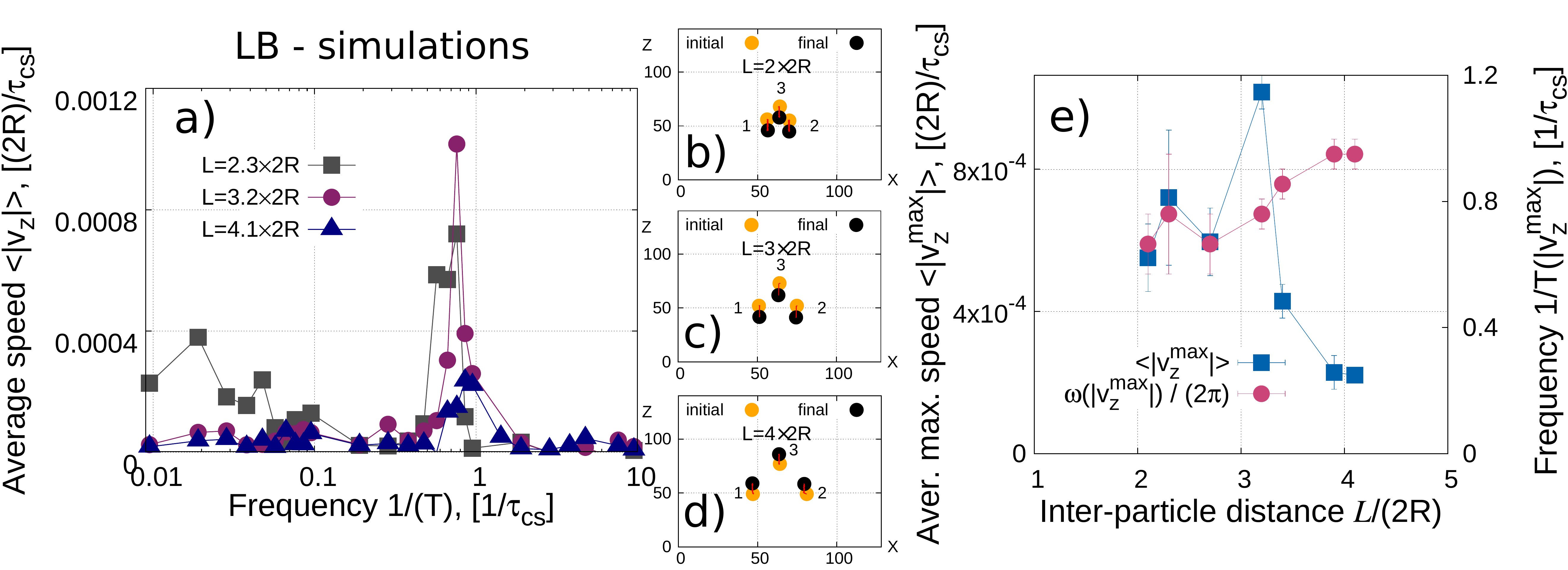}
\caption{a) Average speed of the center of mass of the swimmer as a function of the frequency of the oscillating field illustrated for various values of $L_{ij}$ obtained by LB-simulations. The three panels in the middle (b, c, d) illustrate the motion of the swimmer for different $L_{ij}$. Further parameters: $Bo^{\mathrm{LB}}=0.16$, the time-dependent field is applied along the X-axis (Fig. \ref{fig:geom}) and the field ratio $|B(t)|/|B|\approx 0.36$. e) Maximum average speed and frequency at maximum speed of the swimmer's center of mass as a function of $L_{ij}$ extracted from Fig. \ref{fig:freq}a \rev{and further values of $L_{ij}$. At low $L_{ij}$ uncertainties of the exact positions of the maximum speed are shown resulting from the broadened peaks. Simulations performed for $L_{ij}$ lower than $2\times 2R$ typically result in one or even all the beads sinking, since the changing particle orientation strongly perturbs the highly curved interface which can lead to a detachment of the particle. In contrast, at high $L_{ij}>4\times 2R$, the swimmer speed is so low that no visible motion is observed.}}
\label{fig:freq}
\end{figure*} 

\rev{The most striking feature of our magnetocapillary swimmer is its response to changes in frequency of the external driving force. While not all cyclic driving protocols yield self-propulsion, if we fix the directions of both the external static and the time-dependent field as shown in \figref{geom}, the average speed $\left<v_{\mathrm{z}}\right>$ experiences a sharp peak showing that there is a driving for which the swimmer is most efficient given its geometry. However, the effect of the geometry is non-trivial (Fig. \ref{fig:freq}e) as the average speed of swimming at the optimum driving frequency first increases (for $L_{ij}<3\times 2R$) and then decays (for $L_{ij}>3\times 2R$).}

\rev{A similar non-monotonous frequency response was previously reported in the experimentally realized linear magnetocapillary swimmer \cite{GrHu16}, where, due to the non-Stokesian nature of the device, the swimming speed demonstrates a true resonance at the characteristic frequency of the harmonic oscillator $\tau_{\mathrm{HO}}$ with}

\rev{\begin{equation}
\tau_{\mathrm{HO}}=\sqrt{\frac{m_{\mathrm{R}}}{k}}.
\label{eq_HO}
\end{equation}
Here, $m_{\mathrm{R}}=m/2$ and $m$ is the mass of the bead and $k$ the curvature of the potential between the beads \cite{LaGr16}. Nonetheless, in this case the resonant frequency monotonously decreases with increasing $L_{ij}$. In the case of a triangular geometry, the velocity-frequency relation could not be determined experimentally, while the modelling showed a more complex behaviour as a consequence of the observed inertial periodic rotation of the beads \cite{REFMaxPhD}. However, while an interesting direction to pursue, the peak observed in LB simulations (Fig.~\ref{fig:freq}) cannot be associated with $1/\tau_{\mathrm{HO}}$, and actually appears at up to an order of magnitude smaller frequencies.}   

\rev{Notably, an optimal velocity as a function of frequency was also found in triangular bead-spring swimmers in the pure Stokes regime \cite{RiFa18}. Calculation of the swimming velocity within a perturbative approach to the leading order in $R/L$,  provides the dependency of $v\sim L^{-2}$. This dependency, as well as the existence of the peak velocity in the frequency domain seems to be insensitive to the geometry of the swimmer and the details of the driving protocol. Namely, this behaviour was also found for the linear swimmer, using the same methodology \cite{PaSm15, Feld06, PaMe17, RiFa18}. However, at the same level of theory, the resonant frequency turns out to be independent of the swimmer size \cite{PaSm15, PaMe17, RiFa18}, despite the fact that  our simulations  suggest a more complex behaviour (Fig. \ref{fig:freq}e). }

\rev{Interestingly, when all orders in $R/L$ are taken into account, for a linear swimmer \cite{Feld06}, an increase of the optimum frequency $\omega\ind{St}$ with the inter-particle distance $L$, and $C_1$, $C_2$, and $C_3$ being functions of system parameters\cite{Feld06}
\begin{equation}
\omega_{\mathrm{St}}(L) ~\sim \frac{1}{L}\sqrt{C_1+C_2 L+C_3 L^2},
\label{eq_ana}
\end{equation}
has been reported. This result, obtained in the linear arrangement of beads, seems to capture the general trends as observed in the simulations. }

\rev{To check if such a behavior is expected in a triangular geometry, we numerically calculate the swimming velocity of an externally driven, triangular bead-spring device (Appendix Fig. \ref{im:3}), by this avoiding the truncation of the perturbation series in the orders of $R/L$ (see Appendix for details). The model recovers the $\sim L^{-2}$ dependence of the maximum speed (Appendix Fig.~\ref{fig:freq_BSM}) \cite{RiFa18, PaSm15, Feld06, PaMe17} as well as the dependence of the optimal frequency as a function of the increasing distance between the beads given by Eq.~(\ref{eq_ana}) ($C_1, C_2, C_3$ are here treated as fitting parameters). }

\rev{Using dimensional analysis, however, one can show that the emergent velocity of the swimmer in all Stokes perturbative models depends only on $R/L$,  $A/(k R)$, and $\omega\tau_{St}$, where $A$ denotes the amplitude of the applied forces. Here, $\tau_{\mathrm{St}}$ is the characteristic time scale of relaxation of the swimmer spring  \cite{PaMe17}, relative to the Stokes drag 
 \begin{equation}
\tau_{\mathrm{St}} = \frac{6 \pi \eta R}{k}. 
\label{eq:tv}
\end{equation}
In the absence of inertia, $\tau_{\mathrm{St}}$ sets the internal time scale of the swimmer. Unfortunately, $\tau_{\mathrm{St}}$ is not recovered in LB simulations as the time scale is characteristic for the frequency associated with the peak speed, and appears at up to an order of magnitude larger frequencies. }

\rev{In fact, we find  this peak to be characterized by the time scale of bead coasting through the fluid $\tau_{\mathrm{cs}}$ 
\begin{equation}
\displaystyle \tau_{\mathrm{cs}}=\frac{m}{6\pi\eta R}, 
\label{eq:tauv}
\end{equation}
\rev{This is a characteristic time for a spherical particle placed in a bulk fluid needed to come to equilibrium,  whose equation of motion is governed by Newton's law with a Stokes drag, i.e. $m dv/dt = - 6 \pi \eta R v$}. As it is clear from \eqnref{tauv}, the relaxation time changes either upon variation of the bead radius or the fluid viscosity, hence the properties of the fluid. Indeed, scaling frequencies using $1/\tau_{\mathrm{cs}}$ allows us to collapse all the peaks in Fig.~\ref{fig:freq}a, suggesting that the optimal swimming speed in our LB simulations arises from integrating coasting in most appropriate manner. Consequently, the average maximum speeds are reached for moderate $L_{ij}$ ($\approx 3.2\times 2R$) and are of the order of $\approx 0.001\cdot 2R$ per period of the oscillating field as seen in Fig.~\ref{fig:freq}a. This result is remarkable because it shows  
%, while our LB simulations clearly do not recover motion %dominated by the bead mass observed in experiments, despite %the fact that the the Reynolds number is very small, 
%ALEX+JENS: We could not understand the long sentence and %found it confusing. The LB story is maybe not so important %here.
the emergence of inertial effects which cannot be captured by a simple Stokes dynamics.}

\rev{The Stokes model (Appendix A) is, however,  capable of reproducing several central results of LB-simulations: 
1) The swimmer propagates perpendicularly to the excitation (Figs.~\ref{fig:freq}b-d, and \ref{fig:freq}e); 
2) The average speed of swimming decreases with increasing inter-particle distance in $L>3\times 2R$, which is in agreement with the far-field description of hydrodynamics in the model. For small $L_{ij}$, the LB simulations behave differently due to short-range capillary interactions between the spheres which cannot be described by the bead-spring model. Notably, higher order treatments of the hydrodynamic interactions do not change the trend in the frequency response.
3) The peak in the velocity-frequency response is sensitive to the inter-particle distance.
4) The characteristic frequency grows with increasing $L$. As for the behaviour of the average speed of swimming, the LB-simulations are expected to behave differently for small $L$.} 

\rev{Besides the dependency of speed dependence on the frequency (\figref{freq}a), several additional effects cannot be captured by our Stokes modes. Namely, at small $L_{ij}$ ($\approx 2.3\times 2R$) the swimmer moves efficiently for a wide range of frequencies and the main peak is broadened by the relative proximity of the particles. }

\rev{The motion at low frequencies is, furthermore, dominated by the sizable deformations of the interface and the capillary potential has a much more complex structure than described by Eqs. (\ref{eq_4}) and (\ref{eq_7}), making the determination of the spring constant even more challenging, given the current setup of the simulation. Importantly, the deformations of the interface introduce an additional time scale $\tau_{\mathrm{int}}$
\begin{equation}
\displaystyle \tau_{\mathrm{int}}=\frac{R\eta}{\gamma}, 
\label{eq:tauint}
\end{equation}
Interestingly, using this time scale equally well collapses the peaks shown in Fig. 9a, suggesting that this could be the relevant process dominating the dynamics. However, our test simulations with a significantly larger box do not reproduce this result, while the coasting time scale reappears as the relevant one in the system.}

\section{Summary}

In this study we performed extensive lattice Boltzmann simulations of magnetocapillary swimmers consisting of three ferromagnetic beads trapped at a fluid/fluid interface. The simulations are inspired by the experiments reported in Refs. \cite{VaCl12, LuOb13, GrLa15}, where ferromagnetic beads placed at the water/air interface were driven by magnetic fields and reached propulsion speeds of hundreds of $\mu$m/s.

To match the corresponding regimes for the capillary potentials of the experiments we studied equilibrium properties of a single (Fig.~\ref{fig:1_bead}), two (Fig.~\ref{fig:2_beads}) and three particles (Fig.~\ref{fig:3_beads}) at the interface. The resulting dependence of the center-to-center distance for two and three particles on the strength of the magnetic moment unveiled three main regimes that are in agreement with the experimental observations (Figs. 3 and 6 of Ref. \cite{VaCl12}).

Adding a smaller oscillating magnetic field drives the system out of the local equilibrium and results in a directed motion of the three-bead swimmer (\figref{traj}). By fixing the direction of the oscillating field we demonstrated (\figref{freq}a) that the average speed of the swimmer nonmonotonously  depends on the frequency of the $\vec{B}(t)$-field and reaches its maximum at approximately $0.001$~$2R/\tau^{\mathrm{LB}}_{\mathrm{v}}$ for frequencies in the vicinity of the inverse coasting time defined by \eqnref{tauv}. \rev{The emergence of this time scale is interesting as the swimmer operates in near the Stokes regime, yet the inertial effects start to appear.
Consequently the time scale of the swimmer characteristic for $Re=0$\cite{PaSm15} is not determined as the relevant one, although certain elements of the Stoksian dynamics are recovered.  Furthermore, the motion observed in LB simulations is significantly different to experiments with magnetocapillary swimmers (Refs. \cite{VaCl12, LuOb13, GrLa15}), since in these systems the motion is dominated by inertia and a classical resonance is observed, at least in the linear geometry. The understanding of the emergent time-scale $\tau_{cs}$ requires thus further theoretical investigations which we hope to undertake in future.}

%\rev{In the current work, however, we attempted to reach the experimental regime by increasing the system size. As expected, the obtained trends were consistent with changes in the capillary potential, and the modification of the spring constant of the total magnetocapillary potential (Eq.~\ref{eq_an_1} in the Appendix). The resulting propagation speed could thus increase by a factor of two, however, never reaching speeds observed in the experiment.}

The LB-simulations performed here also point out to the importance of simultaneous optimization of the swimmer's geometry and driving which is a very important property for applications: swimmers with different inter-particle distances ($L_{ij}$) might be controlled independently by application of AC-fields with distinct frequencies. In the context of the current work, however, besides identifying these interesting phenomena, we clearly establish lattice Boltzmann simulations as a technique for studying microswimming on liquid interfaces, which hitherto was not possible.

\rev{\subsection{Appendix: Triangular bead-spring swimmer in the Stokes regime}}

\begin{figure}[htb]
\centering
\includegraphics[scale=0.4]{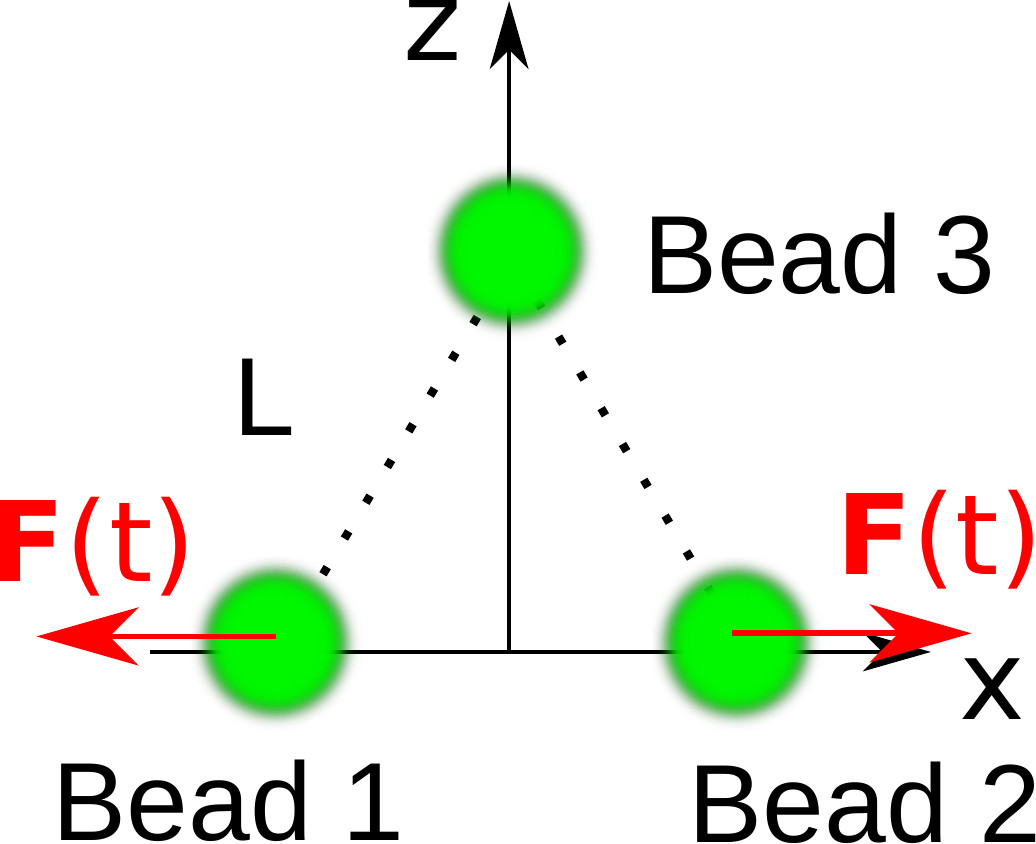}
\caption{Sketch of the initial configuration of the swimmer modeled using the Felderhof approach.}
\label{im:3}
\end{figure}

We assume three beads of equal radii $R$ pairwise connected by harmonic springs of spring constant $k$ and equilibrium length $L$ (Fig. \ref{im:3}), such that the potential energy of a spring connecting two beads at positions $\vec{r}_i$ and $\vec{r}_j$ reads 
\begin{equation}
\phi_{ij} \equiv \phi (\vec{r}_i - \vec{r}_j) = \frac{1}{2} k (|\vec{r}_i - \vec{r}_j| - L)^2. 
\label{eq_an_1}
\end{equation}

In addition to the spring constant $k$, external forces denoted by $\vec{F}_i (t)$ act on each bead $i$, satisfying the condition that (a) the sum over all forces $\vec{F}\ind{tot}(t)=\sum_{i=1}^3 \vec{F}_i(t)=0$ as well as (b) the resulting torque $\vec{T}\ind{tot}(t)=\sum_{i=1}^3 \vec{r}_i \times \vec{F}_i(t)=0$ both vanish at each time step. Note that the origin of the coordinate system relative to which the torque is calculated is irrelevant when (a) holds.  For the calculations below, we specifically set the swimmer  in the $XZ$-plane, and exert a sinusoidal force specified by $\vec{F}_1(t) = A \sin(\omega t) \vec{e}\ind{x}$ on the first bead. Consequently, forces on the other two beads  are set by the constraints (a) and (b). Notably, this protocol allows only for pure translations of the swimmer's center of mass in the steady state.   

In the regime of the bead radius $R \ll L$, the hydrodynamic interaction between the beads is given by the Oseen tensor \cite{Osee10} 
\begin{equation}
\!\!\vec{\hat{T}}(\vec{r}_j\!-\!\vec{r}_k)\!=\! \frac{1}{8 \pi \eta |\vec{r}_j-\vec{r}_k|} \left(\!\vec{\hat{I}}\!+\!\frac{(\vec{r}_j - \vec{r}_k) \otimes (\vec{r}_j - \vec{r}_k)}{(\vec{r}_j-\vec{r}_k)^2}\!\right)
\label{eq_an_3}
\end{equation}
and the equation of motion for each bead is governed by the Stokes law, such that the system is described by the set of equations of motion describing the time evolution of the position $\vec{r}_i$ of each bead
\begin{equation}
\displaystyle 
\begin{split}
\frac{d \vec{r}_j}{dt} & = \nu_j \left(\vec{F}_j(t) + \sum_{k \neq j} \vec{G}(\vec{r}_j - \vec{r}_k)\right) \\ 
& + \sum_{k \neq j} \vec{\hat{T}}(\vec{r}_j - \vec{r}_k) \left(\vec{F}_k(t) + \sum_{l \neq k} \vec{G}(\vec{r}_k - \vec{r}_l)\right),
\end{split}
\label{eq:Felderhof}
\end{equation}
where $\nu_j = \frac{1}{6 \pi \eta R}$ and $\vec{G}$ is the spring force $\vec{G}(\vec{r}) := - \vec{\nabla} \phi(\vec{r})$. The first (mobility) term of Eq. (\ref{eq:Felderhof}) describes the motion due to external and spring forces directly acting on the beads, while the second accounts for the hydrodynamic interactions between the beads.

The swimming velocity emerges from inserting the solution of Eq. (\ref{eq:Felderhof}) into Eq. (\ref{eq:v}). With the above-prescribed driving, the swimmer's motion is found always to be associated with a translation in the $Z$-direction without a net rotation. 

\begin{figure*}[htb]
\centering
\includegraphics[width= 0.9\textwidth]{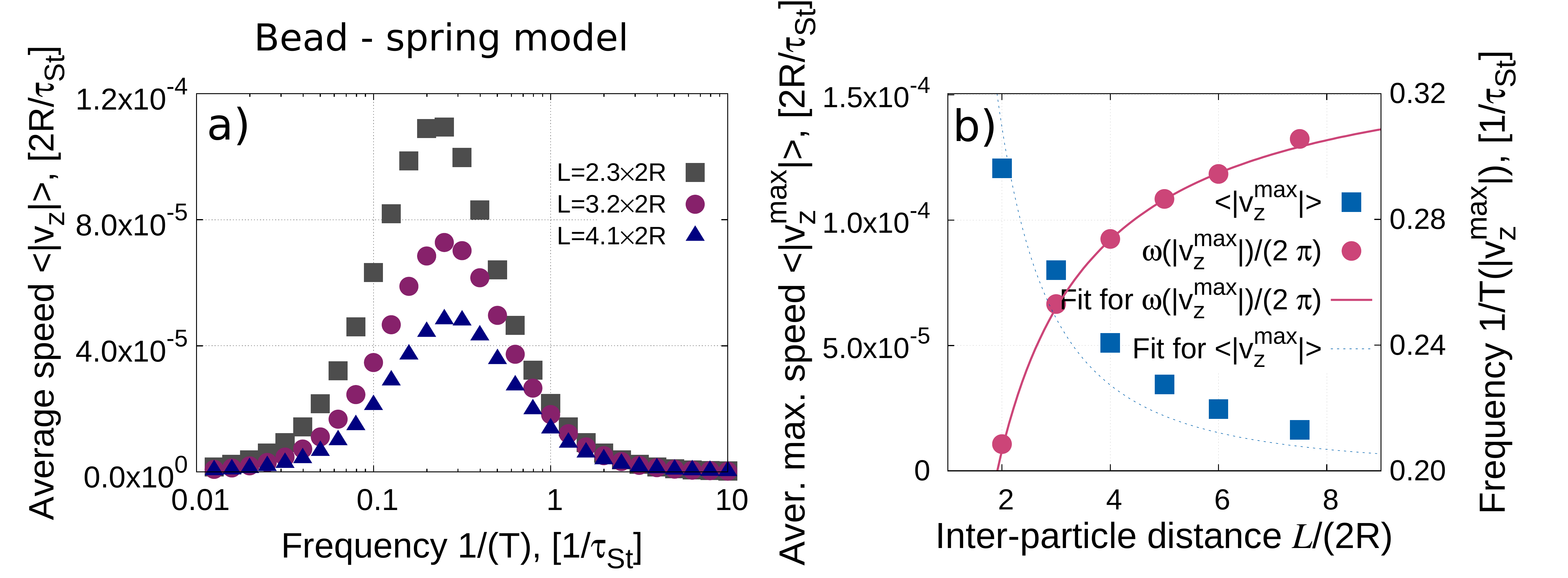}
\caption{a) Average speed of the center of mass of the triangular swimmer (Fig.~\ref{im:3}) on the driving frequency obtained numerically within the bead-spring model (Eq. (\ref{eq:Felderhof})). Parameter used: $A/(k R) = 1.0$. b) \rev{Maximum speed (blue squares) and frequency corresponding to the maximum speed} of the swimmer (pink dots) determined from the bead-spring model. Fits are made using $|v\ind{z}| \ind{max}(L)=A_1/L^2$ (average speed, $A_1$ a fitting parameter) and Eq. \eqref{eq_ana}.
}
\label{fig:freq_BSM}
\end{figure*} 
~\\

\begin{acknowledgments}
This work was financially supported by the DFG Priority Programme SPP 1726 ``Microswimmers – From Single Particle Motion to Collective Behaviour'' and the Cluster of Excellence ``Engineering of Advanced Materials'' (project EXC 315). We further acknowledge the J\"{u}lich Supercomputing Centre (JSC) and the High Performance Computing Center Stuttgart (HLRS) for the allocation of computing time.
\end{acknowledgments}

\bibliography{bibliography_MCS}% Produces the bibliography via BibTeX.

\end{document}